\documentclass[lettersize,journal]{IEEEtran}

\usepackage[nolist,nohyperlinks]{acronym}
\usepackage{amsmath,amsfonts,amssymb}
\usepackage{algorithmic}
\usepackage{algorithm}
\usepackage{array}
\usepackage{authblk}

\usepackage{bbm}

\usepackage{dsfont}

\usepackage[caption=false,font=normalsize,labelfont=sf,textfont=sf]{subfig}
\usepackage{textcomp}
\usepackage{stfloats}
\usepackage{url}
\usepackage{verbatim}
\usepackage{cite}

\usepackage{graphicx} 
\usepackage[textwidth=16cm,textheight=21.7cm]{geometry}

\usepackage{physics}

\usepackage{todonotes}

\newacro{CNOT}{controlled $X$}
\newacro{CPTP}{Completely-Positive Trace-Preserving}
\newacro{GHZ}{Greenberger-Horne-Zeillinger}
\newacro{MSE}{Mean-squared error}
\newacro{NISQ}{Noisy Intermediate-Scale Quantum Era}

\newacro{QCRB}{Quantum Cramèr-Rao Bound}
\newacro{QEC}{Quantum Error Correction}
\newacro{QFI}{Quantum Fisher Information}
\newacro{QFIM}{Quantum Fisher Information Matrix}
\newacro{QKD}{Quantum Key Distribution}
\newacro{QNT}{Quantum Network Tomography}
\newacro{QPT}{Quantum Process Tomography}
\newacro{QST}{Quantum State Tomography}

\title{Quantum Network Tomography}

\author[1]{Matheus Guedes de Andrade}
\author[2]{Jake Navas}
\author[3]{Saikat Guha}
\author[2]{Inès Montaño}
\author[4]{Michael Raymer}
\author[4]{Brian Smith}
\author[1]{Don Towsley}

\affil[1]{Manning College of Information and Computer Science, University of Massachusetts Amherst}
\affil[2]{Department of Applied Physics and Materials Science, Northern Arizona University}
\affil[3]{Wyant College of Optical Sciences, University of Arizona}
\affil[4]{Department of Physics, University of Oregon}

\date{May 2023}

\begin{document}

\maketitle

\begin{abstract}
    Errors are the fundamental barrier to the development of quantum systems. Quantum networks are complex systems formed by the interconnection of multiple components and suffer from error accumulation. Characterizing errors introduced by quantum network components becomes a fundamental task to overcome their depleting effects in quantum communication. Quantum Network Tomography (QNT) addresses end-to-end characterization of link errors in quantum networks. It is a tool for building error-aware applications, network management, and system validation. We provide an overview of QNT and its initial results for characterizing quantum star networks. We apply a previously defined QNT protocol for estimating bit-flip channels to estimate depolarizing channels. We analyze the performance of our estimators numerically by assessing the Quantum Cramèr-Rao Bound (QCRB) and the Mean Square Error (MSE) in the finite sample regime. Finally, we provide a discussion on current challenges in the field of QNT and elicit exciting research directions for future investigation.
\end{abstract}
\section{Introduction}

The promise of engineering useful quantum systems that will revolutionize communications is limited by the ability to correct errors and mitigate noise~\cite{eisert2020quantum}. This fundamental obstacle is pervasive across the quantum network architecture: quantum networks are formed by the interconnection of multiple quantum systems, suffering from multiple sources of error, which includes both external noise and system imperfections. Thus, characterizing network errors is the initial step to ameliorate its impact in quantum communication. Characterization is fundamental since knowledge on errors improves the efficiency of error decoders and purification protocols, and provides information for error-aware quantum network protocols, e.g., fidelity-aware entanglement routing, and network management, e.g., faulty hardware identification.

\ac{QNT} is centered on error characterization, merging ideas from classical network tomography and quantum tomography to provide end-to-end error characterization of quantum networks~\cite{de2022quantum, de2023characterization}.
It adds to the literature of quantum error characterization~\cite{flammia2020efficient, wagner2022pauli} by focusing on the problem of quantum channel estimation when channels cannot be directly used for estimation.

End-to-end characterization of internal network components has proven valuable to the development of the Internet, enabling the evaluation of network performance without relying on administrative access~\cite{castro2004network, adams2000use}. It is reasonable to expect the quantum Internet to follow the structure of its classical counterpart as it matures, emerging from the connection of multiple quantum networks managed by different entities. Hence, \ac{QNT} becomes relevant as it provides tools for characterizing internal quantum network behavior from end-to-end measurements.

Error characterization is also relevant to \ac{NISQ} devices~\cite{preskill2018quantum}. \ac{QNT} provides ways to validate network protocols and design, since end-to-end characterization can be used together with direct channel characterization to test and debug entanglement distribution protocols. These protocols are essential to both computing~\cite{cacciapuoti2019quantum} and sensing applications, being the key services provided by quantum networks. 

Classical network tomography protocols estimate performance metrics, e.g., link loss and delays, by performing end-to-end traffic measurements. Both in unicast and multicast tomography, users exchange messages that allow for indirect measurements of internal network properties~\cite{he2021network}. Measurements are indirect since the observed end-to-end behavior is a composition of the behavior of individual links used for communication, e.g., the sum of link delays. Protocols then explore correlations on measurements performed at different nodes to infer individual parameters. Similarly, \ac{QNT} relies on quantum state distribution---users utilize the network to prepare a remote quantum state using the network to propagate entanglement---to infer link error characteristics such as Pauli error probabilities. In this context, measurement statistics of distributed states depend on errors introduced by each link used for communication and can be used for estimation.

The differences between classical and quantum information introduce new caveats to the study of the network tomography problem. Most metrics studied in the classical case compose in an additive way, while metrics of interest in quantum networks often do not. For instance, the composition of two bit-flip channels does not yield a composed bit-flip channel with probability given by the sum (or product) of the probabilities of the two individual channels. Additionally, QNT protocols can select between different quantum states to distribute, which encode parameters differently, and perform different measurements to extract parameter information for estimation.
Despite these differences, it is possible to draw a parallel between classical unicast communication with bipartite quantum state distribution, and classical multicast communication with multipartite quantum state distribution. In recent work, we have explored the latter connection to lay down an initial formulation for \ac{QNT} based on multipartite entanglement distribution~\cite{de2022quantum}. Our goal with this article is to provide a high-level overview on the recent efforts towards the characterization of quantum networks with QNT and to push the state-of-the-art by providing results on the application of a previously defined \ac{QNT} protocol in a novel setup.
Our contributions are as follows:
\begin{itemize}
    \item We provide an overview of current \ac{QNT} methods expressed within the framework defined in~\cite{de2023characterization}, highlighting critical assumptions to their initial formulation.
    \item We illustrate \ac{QNT} methods by adapting a protocol initially introduced in~\cite{de2022quantum} for a star network whose links are modeled as bit-flip channels to characterize stars with depolarizing channels. We evaluate this method by computing the Quantum Cramèr-Rao Bound and assessing the performance of estimators in the finite sample regime through simulation. The novelty of this contribution lies in the characterization of networks with depolarizing channels, showcasing the power of the QNT framework. 
    \item We elicit the main challenges and obstacles to the development of \ac{QNT} protocols capable of characterizing arbitrary quantum networks, and propose new research directions in \ac{QNT}.
\end{itemize}

From now on, we refer to star quantum networks with bit-flip and depolarizing links as flip and depolarizing stars, respectively.

\section{Quantum Network Tomography}

The overarching goal of \ac{QNT} is to characterize errors introduced by different network components through descriptions of their underlying \ac{CPTP} maps through end-to-end measurements. End-to-end measurements refer, in this case, to quantum measurements performed by end-nodes on states distributed through the network.

\subsection{The QNT Framework}

\begin{figure*}
    \centering
    \includegraphics[scale=0.52]{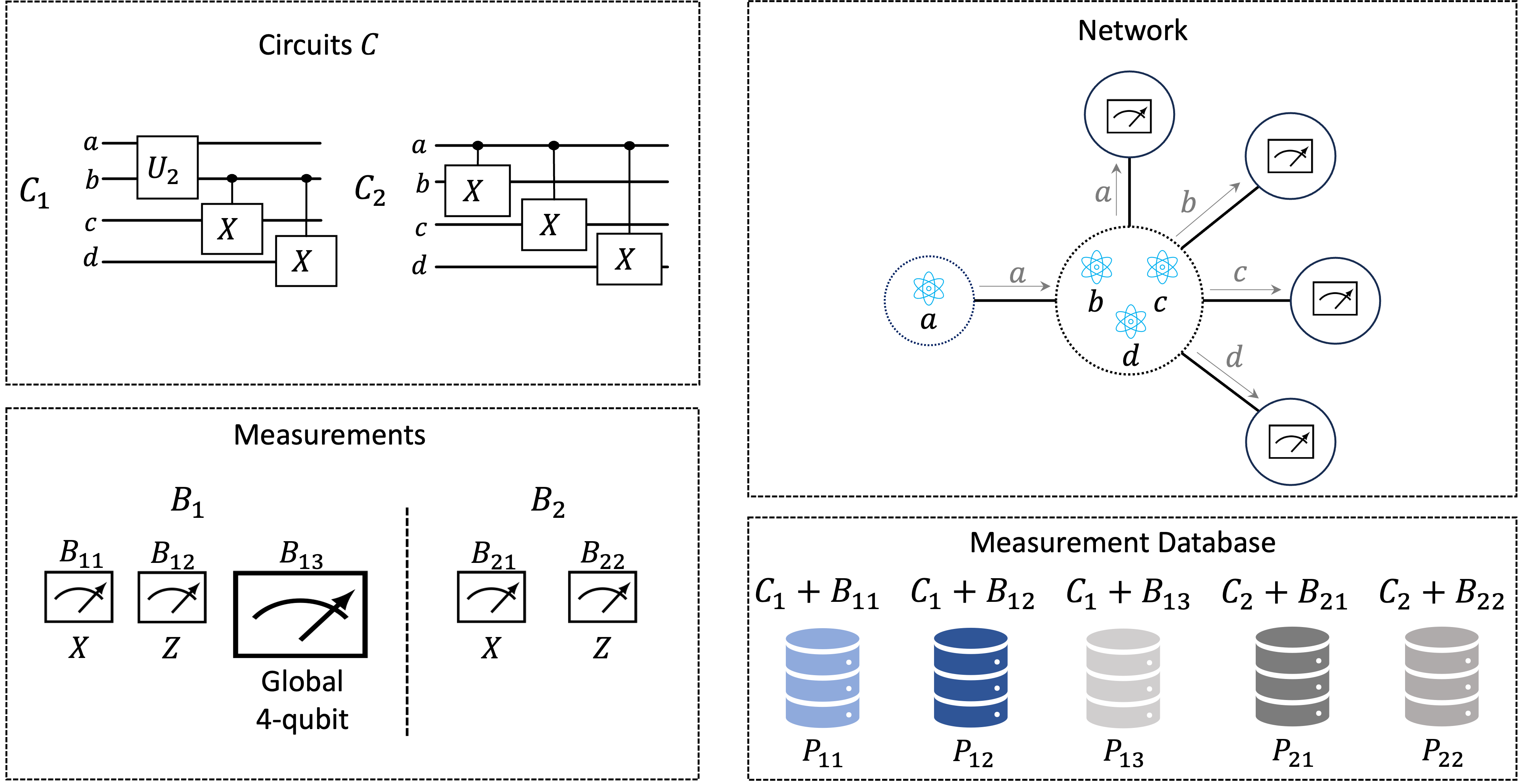}
    \caption{\ac{QNT} framework example. Two distribution circuits $C_1$ and $C_2$ are used to generate a measurement database with five components. In this example, $U_2$ represents an arbitrary 2-qubit gate. We show how qubits are placed in the star through atoms with indices matching the circuit description. Arrows indicate which qubits are transmitted through the links. Each copy of a state distributed through $C_1$ is measured either with local measurements in the $X$ basis, local measurements in the $Z$ basis, or a global measurement in a $4$-qubit entangled basis, e.g., . For $C_2$, local measurements in the $X$ and $Y$ basis are used. Each component (colored disks) of the measurement database corresponds to the combination of a state distribution circuit with a measurement operator.}
    \label{fig:framework}
\end{figure*}

Quantum state distribution is the basis for \ac{QNT} in both one- and two-way quantum communication networks, and having a unified framework to describe \ac{QNT} protocols based on state distribution is fundamental.
There are different models of quantum state distribution one can assume when considering the description of tomography protocols. A general framework for QNT must apply regardless of the state distribution model provided by the network. As an example, a network can strictly provide EPR pair generation between any two end-nodes, and estimation has to rely exclusively on bipartite entanglement.
In previous work~\cite{de2023characterization}, we proposed a framework based on state distribution that is applicable to arbitrary QNT problems. We now describe this framework, which is depicted in Fig.~\ref{fig:framework}:
\begin{enumerate}
    \item Select a set of $N$ distribution circuits $C = \{C_0,\ldots, C_{N - 1}\}$. Here a quantum distribution circuit is a combination of network links and quantum logic operations in the nodes used to distribute copies of a quantum state. Note that a distribution circuit is equivalent to a state distribution service provided by the network.
    \item For each circuit $C_j \in C$, define a set $B_{j}$ of $M_j$ measurement bases, $j \in \{0,\ldots, N - 1\}$.
    \item For each combination of circuit $C_j \in C$ and basis $B_{jk} \in B_{j}$, ,  define a number $P_{jk}$ of probes, i.e., of distributed states, $k \in \{0,\ldots, M_{j} - 1\}$.
    \item Create a dataset with $\sum_{jk}P_{jk}$ of measurement observations for estimation.
\end{enumerate}
In essence, this framework translates \ac{QPT} to the network scenario, providing enough flexibility to users to determine the distribution circuits used---constrained by the state distribution services provided by the network---as well as the quantum measurements performed in the end-nodes.

\subsection{Evaluating estimation performance}\label{sec:estimationPerformance}

The \ac{QNT} framework allows for multiple protocols that can characterize network noise, and understanding their differences in terms of estimation efficiency is key to the practical application of \ac{QNT}. Focus has been on analyzing estimator performance in terms of the \ac{QFIM} and the \ac{QCRB}. States distributed within the framework are described by density matrices $\rho(\vec{\theta})$ that depend on the link parameters $\vec{\theta}$. The \ac{QFIM} of a quantum state $\rho(\vec{\theta})$ quantifies how much information from $\vec{\theta}$ is captured in a single copy of the state. Moreover, the \ac{QFIM}'s inverse of $\rho(\vec{\theta})$ provides a lower bound on the covariance matrix of parameter estimates for $\vec{\theta}$ based on state measurements, which is known as the \ac{QCRB}. The QFIM relates to the maximum information on a parameter that can be extracted from a quantum system, and the \ac{QCRB} provides fundamental limits for the efficiency of estimators. In this context, the optimal \ac{QNT} protocol for network characterization in terms of variance is the one that maximizes the QFIM. Despite this straightforward characterization of optimallity, using optimal estimators in terms of the QCRB may require expensive non-local measurements, which demand the consumption of pre-shared entanglement. Therefore, there is an interesting trade-off between estimation efficiency and network resource utilization which \ac{QNT} protocols can explore. 

There also exist intrinsic difficulties in finding optimal \ac{QNT} protocols by maximizing the \ac{QFIM}. Computing the matrix for an arbitrary state $\rho(\theta)$ is hard, both theoretically and computationally. Finding closed form expressions for the \ac{QFIM} is highly non-trivial, and computing its value numerically has exponential time complexity with the size of the state. Finding protocols that maximize the QFIM has the additional difficulty of solving an optimization problem on the space of all possible quantum state distribution circuits, which is intractable in the general case. Despite these difficulties, we focus on numerical evaluations of the QFIM for small systems to probe the efficiency of our proposed methods.

\subsection{Assumptions}\label{sec:assumptions}
In the most general case, a complete, end-to-end characterization of network errors is practically infeasible in large networks without additional assumptions. The reasons are two-fold. First, there are multiple  errors corrupting quantum states being transmitted across a network, such as memory decoherence and gate and measurement errors. These errors accumulate as network links and nodes are used for communication. Providing a complete description of the \ac{CPTP} maps via end-to-end measurements for each of these error sources becomes impractical, if not impossible.
Second, \ac{QST} and \ac{QPT} have exponential sample complexity in state and channel dimension, respectively, and are impractical for network characterization without additional assumptions.
For instance, suppose that multiple network users want to characterize internal network noise through state distribution with \ac{QST}. They must first characterize the states being distributed by the network, which requires an exponential number of samples in the dimension of the state. Then, parameters must be estimated from the state description, that is itself exponential in the dimension of the Hilbert space, in the worst case. Therefore, we focus on a \ac{QNT} formulation that operates under the following assumptions to enable practical end-to-end characterization of internal network noise:
\begin{enumerate}
    \item \textbf{\textit{Error-free quantum gates and memories.}} It is challenging to obtain a separate description for noise introduced by each network component. Therefore, we assume perfect quantum operations and memories in the nodes and focus on estimating noise introduced by network links.
    \item \textbf{\textit{Estimation of Pauli errors.}} Each link $l$ is assumed to represent a single-qubit Pauli channel $$\mathcal{E}_l(\rho) = \sum_{k \in \{\mathds{1}, X, Y, Z\}} \theta_{lk} \sigma_k \rho \sigma_{lk}$$ determined by a three-dimension parameter vector $\vec{\theta}_l$ to be estimated, where each $\sigma_{k}$ denotes a Pauli operator. In one-way quantum communication, the Pauli channel $\mathcal{E}_l$ corresponding to link $l$ directly transforms any qubit sent through the channel. In two-way, states generated through link $l$ are mixed states diagonal on the Bell basis that emerge from the application of $\mathcal{E}_l$ to one of the qubits of a pure Bell state. In this paper, we present solutions for the characterization of depolarizing channels, which are a particular case of Pauli noise characterized by a single parameter.
    \item \textbf{\textit{Quantum Circuit Distribution Requests.}}
    Users can request a particular distribution circuit to be used in each intermediate node, although intermediate nodes cannot contribute with measurement data for parameter estimation. This assumption maps to networks that provide a flexible state distribution service where users can determine the distribution circuits in exquisite details.
    \item \textbf{\textit{Route selection for Distribution.}} End nodes can specify which intermediates nodes are used to serve a given request.
    \item \textbf{\textit{Parameter estimation with post-selection.}} We analyze estimators without considering channel losses, i.e., we use successfully distributed states for parameter estimation.
\end{enumerate}

These assumptions cast \ac{QNT} into end-to-end quantum parameter estimation, rendering a solution to the problem tractable while maintaining practical and theoretical interest. Focusing on link characterization gives a direct way to quantify how using a particular network link for serving network requests can degrade performance. In addition, it is possible to approximate operational noise in the nodes by pushing its effects to the links under specific conditions, e.g., if both links and operations in the nodes induce Pauli channels.
The assumption of single-qubit Pauli errors provides tractability, since each link requires three parameters to be characterized,
and generality, given that any quantum channel can be projected into a Pauli channel through twirling (random operations applied to the qubit before and after channel action)~\cite{flammia2020efficient}.
Performing \ac{QNT} under the assumption that quantum circuit distribution is specified to the level of granularity of individual node operations and route selection sheds light on the limitations of end-to-end parameter estimation, since it is the most flexible model of entanglement distribution.
Disregarding photonic loss in communication through post selection removes the need for analyzing how qubit losses affect distributed quantum states with the cost of increasing parameter estimation latency, i.e., the time it takes to obtain reliable parameter estimates.

This constrained version of a \ac{QNT} serves as a stepping stone towards more general formulations and we revisit our assumptions in Section~\ref{sec:challenges}, eliciting their limitations and discussing ways to extend \ac{QNT} to broader scenarios.

\subsection{Quantum Flip Star Networks}

Star networks are interesting due to their simplicity and practicality. It is likely that the first operating quantum networks will be stars, and initial quantum network experiments in two-hop networks serve as real-world examples~\cite{hermans2022qubit}. Moreover, we will describe how the links in an arbitrary network can be characterized using techniques applicable to star networks in Section~\ref{sec:arbitraryNets}.
We applied the above \ac{QNT} framework to generate multiple protocols for the characterization of quantum flip stars. Current \ac{QNT} protocols~\cite{de2022quantum, de2023characterization} characterize star networks with Pauli channels described by a single Pauli operator, e.g., quantum bit-flip channels, which we refer to as quantum flip stars. The protocols are based on the six assumptions described, with the additional assumption that each link $l$ is characterized by one parameter $\theta_l$.
They are capable of \textit{identifying} link parameters, i.e. finding a unique solution $\hat{\theta}_l$ for every link $l$ from multiple end-to-end observations, with polynomial sample complexity. Interestingly, the performance analysis carried out in~\cite{de2023characterization} shows that estimation efficiency, in terms of estimation variance and the \ac{QCRB}, depends on parameter values. This indicates that adaptive estimation strategies---\ac{QNT} protocols that dynamically change state distribution based on current parameter estimates---can perform better than static protocols in practical settings. The existing tomography protocols are described in the one-way setting, although can be modified to two-way communication.

Despite the connection between the framework and \ac{QPT}, our protocols differ from the latter on how the measurement dataset is used for estimation. In its vanilla version, \ac{QPT} uses the entire joint distribution of measurements to reconstruct the description of a quantum channel. Our estimators rely, instead, on marginal distributions of qubit measurements, to reduce complexity and enable practical parameter estimation. We used single-~and~two-qubit measurement marginals to identify bit-flip noise in the star, yielding polynomial time estimation in the number of links.

\section{Tomography of Depolarizing Stars}

We now apply the framework to characterize star networks with depolarizing noise. We use the Multicast protocol, initially defined in~\cite{de2022quantum} for the characterization of bit-flip channels, to identify depolarizing parameters in star networks. The single-qubit depolarizing channel is a Pauli channel characterized by one parameter, although with non-zero components for the non-trivial Pauli operators. A depolarizing channel $\mathcal{E}_\theta$ has the form
$$\mathcal{E}_\theta(\rho) = (1 - \theta) \rho + \frac{\theta}{3}(X\rho X + Y \rho Y + Z \rho Z),$$
where $X$, $Y$, and $Z$ denote the Pauli matrices~\cite{nielsen2010quantum}.

\begin{figure}
    \centering
    \includegraphics[scale=0.6]{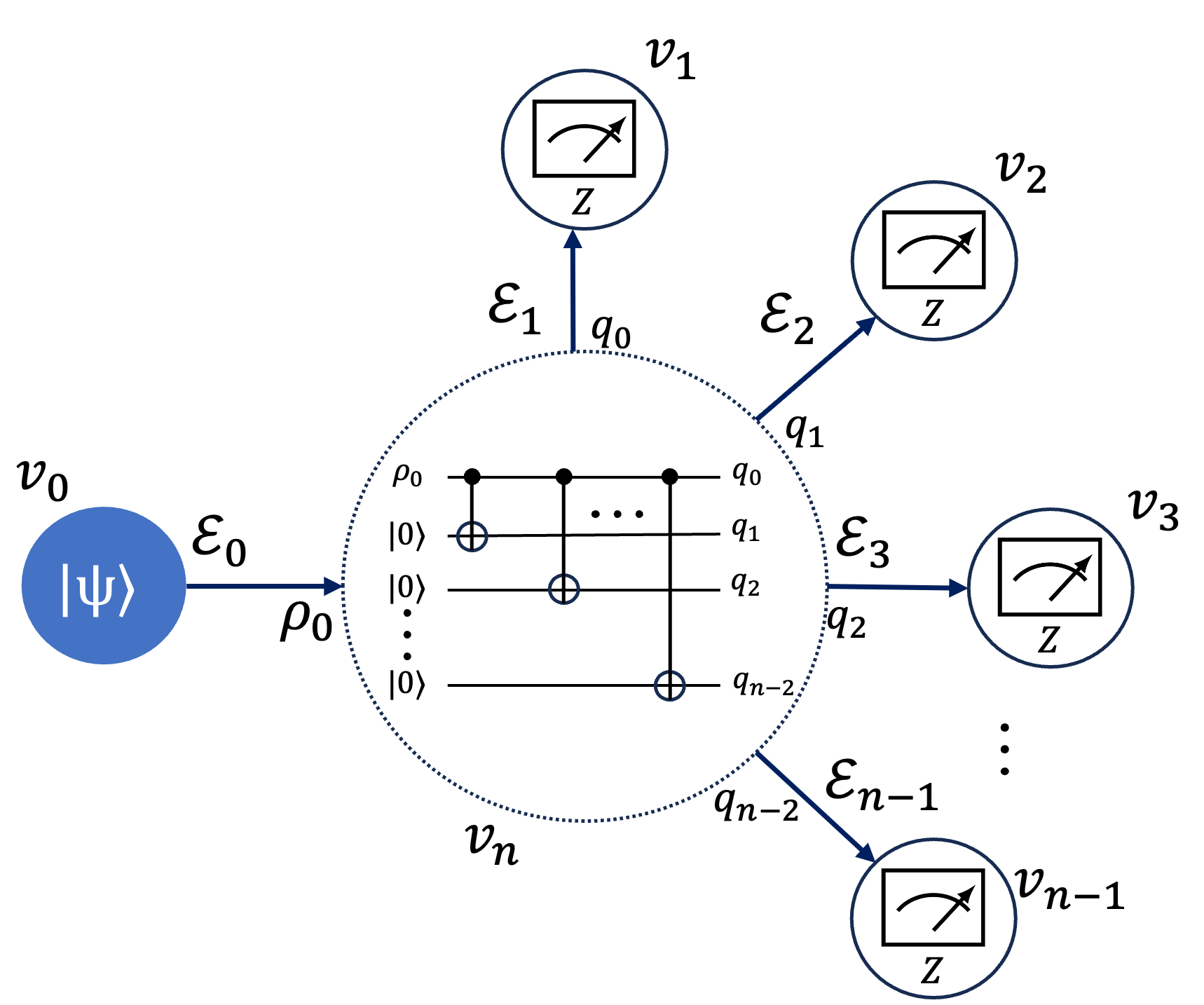}
    \caption{Multicast circuit. Root ($v_0$), intermediate node ($v_{n}$), and leaves ($v_{1}$ to $v_{n-1}$) are depicted as the shaded, dashed line, and continuous line nodes, respectively. Qubits resulting from the multicast circuit, depicted in the figure as $q_i$ for $i \in \{0, 1,\ldots, n - 2\}$, are forwarded to the leaves. The state $\ket{\psi}$ represents either the single-qubit state $\ket{0}$ or the Bell pair $(\ket{00} + \ket{11}) / \sqrt{2}$.}
    \label{fig:multicast}
\end{figure}

The Multicast distribution circuit for a star network of $n$ end-nodes is depicted in Fig.~\ref{fig:multicast} and works as follows. We select an end-node to start the state distribution protocol. From now on, we refer to this node as \textit{root} and to the remaining end-nodes as \textit{leaves}. We apply the Multicast protocol using two initial states for the qubits prepared in the root: a single qubit in state $\ket{0}$ and two qubits prepared in the Bell pair $(\ket{00} + \ket{11}) / \sqrt{2}$. When a single qubit is initialized in the root it is sent to the intermediate node through the interconnecting link. In the case of a Bell pair, one qubit is sent through the link to the intermediate node and the other qubit remains stored in the root. For both cases, the qubit that arrives at the intermediate node is used as the control qubit of $(n - 2)$ CNOT operations targeting $(n - 2)$ newly initialized qubits each in state $\ket{0}$. After the CNOTs are completed, each output qubit is sent to one of the leaves of the star excluding the root. When a qubit in state $\ket{0}$ is used to start distribution, a complete execution of the Multicast protocol yields an $(n - 1)$ qubit state stored in the leaves of the star that is diagonal on the $Z$ basis. When the Bell pair is utilized, an $n$ qubit state diagonal on the $n$-qubit GHZ basis is created in the leaves and the root of the star.

The two initial states utilized for the Multicast protocol are chosen since they allow for the derivation of valid estimators for the depolarizing probabilities of all network links. They also enable an initial analysis on the advantages of using entanglement for estimation. There are many other initial states and protocols that can be used to characterize depolarizing noise in this context, although we do not consider them in this paper.

\subsection{Estimating depolarizing noise with $Z$ diagonal states}

The action of a depolarizing channel of parameter $\theta$ on a qubit prepared on the pure state $\ket{0}$ is equivalent to the action of a bit-flip channel with parameter $\theta_f = (1 - 2\theta / 3)$ applied on the same state. This can be derived by noting that a depolarizing channel randomly transforms $\ket{0}$ to either itself or $\ket{1}$ up to a global phase, and the probability $2\theta / 3$ of observing a flip is that of applying either $X$ or $Y$. Such an equivalence allows us to directly apply the Multicast protocol to characterize the depolarizing probabilities in a star network by using the estimators defined in~\cite{de2022quantum} to compute $\theta_f$ and estimate $\theta$ from it. Placing this in the context of our framework, we have exactly one distribution circuit and one measurement basis. More precisely, the set $C$ contains only the Multicast circuit, such that $N = 1$, and $B$ only contains measurements in the $Z$ basis, yielding $M_{1} = 1$. Hence, we construct the dataset for estimation entirely from Multicast probes measured in the $Z$ basis.
Bit-flip estimates with Multicast probes were shown to have a two-fold degeneracy in~\cite{de2022quantum}. In this case, two estimates for the parameter vector are obtained from a measurement dataset constructed with the Multicast circuit and measurements on the $Z$ basis.
Interestingly, this single step protocol for the depolarizing channel only suffers from the degeneracy of bit-flip estimators for a particular range of parameter regimes. In the bit-flip case, the equations used for estimation have two solutions of forms $\hat{\theta}_f$ and $(1 - \hat{\theta}_f)$. A valid depolarizing probability requires $\theta_f \geq 1 / 3$ and the two-fold degeneracy occurs only when $1/3 \leq \hat{\theta}_f \leq 2 / 3$.


\subsection{Estimating depolarizing noise with GHZ diagonal states}

In order to obtain estimators that utilize distributed GHZ states we
combine equations used in the previous case with the probability of measuring GHZ states with a minus sign, e.g., $(\ket{000} - \ket{111}) / \sqrt{2}$. Measurements of an $n$-qubit GHZ state can be understood as follows. The outcome of the measurement is a bit $b$ and an $(n - 1)$ bit string $s$. The bit $b$ codifies when a minus sign is measured, while the bit string $s$ determines the pattern of bits inside each state in the superposition. For instance, a measurement outcome of $b = 1$ and $s = 01$ implies that the state $(\ket{001} - \ket{110}) / \sqrt{2}$ was measured. The statistics of measurements for $s$ follow the bit-flip statistics for the previous case of a separable state, while the statistics of $b$ provide an additional equation to be used for estimation. Using the statistics for $s$, one can write the equation for the probability of $b = 1$ as single-variable polynomial with degree given by the number of end-nodes in the star. Estimation is complete by computing the root of the $n$-degree polynomial equation, which is an estimate for the depolarizing parameter of the first link, and using it to compute the depolarizing parameters for the different links. 

\subsection{Estimation performance}

We assess the performance of the defined estimation processes analyzing the \ac{QCRB} for the distributed states and the \ac{MSE} for our estimators. For simplicity, we focus on networks where every channel is characterized by the same probability ($\theta_l = \theta^{*}$ for all $l$) following experiments reported in~\cite{de2023characterization}.

\subsubsection{\ac{QCRB}}
\begin{figure*}
    \centering
    \includegraphics[scale=0.48]{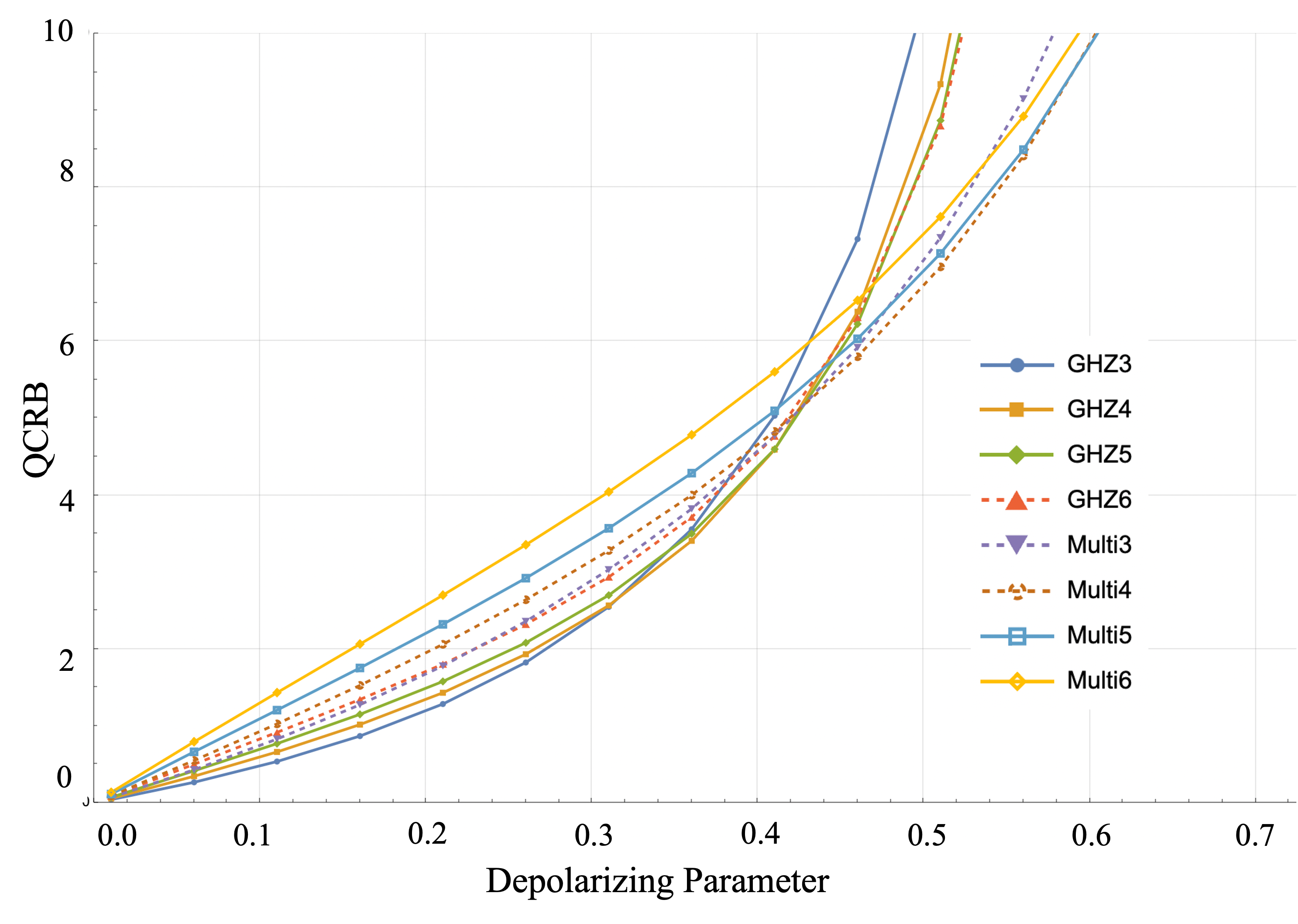}
    \caption{\ac{QCRB}, i.e., QFIM's inverse trace, per depolarizing parameter for $Z$- and GHZ-diagonal states. Curves show the \ac{QCRB} when the Multicast protocol is used to characterize depolarizing stars with different sizes. We investigate the scenario where every channel in the network has the same depolarizing probability. There is a singularity in the estimators corresponding to the case when the depolarizing parameter is 0.75. In this regime, every link in the star outputs the maximally mixed state independent of the state used as input.}
    \label{fig:qfim}
\end{figure*}

As discussed in Section~\ref{sec:estimationPerformance}, the QCRB is a lower bound on the covariance matrix of any estimator for a parameter vector $\theta$ derived from a quantum state $\rho(\theta)$. Fig.~\ref{fig:qfim} shows the \ac{QCRB} with the depolarizing parameter $\theta^{*}$ for star networks with size varying from four to seven for both initial states. We focus on the regime $\theta^{*} \leq 0.75$ since the depolarizing channel always returns a maximally mixed state when $\theta^{*}=0.75$ and the QCRB approaches infinity. The curves show an interesting behavior: there are multiple transition points where the \ac{QCRB} of larger networks become lower than those of smaller ones, which is a trend for both initial states. In particular, once the initial state is fixed, a size four star network has the smallest \ac{QCRB} when the depolarizing probability is small (below 0.3) and the highest \ac{QCRB} when the depolarizing probability is large. This dependence of the \ac{QCRB} on the depolarizing parameter indicates a counter-intuitive result that, for some parameter regimes, estimating large networks is easier than smaller ones. \ac{QCRB}s for large depolarizing probabilities (greater than 0.4) also seem to converge as the number of nodes in the star increases when the initial states are fixed.
Moreover, the relationship between curves show that, in principle, estimators obtained from GHZ-diagonal states can have smaller variance than those obtained with $Z$-diagonal states in the regime of low depolarizing parameter.

\subsubsection{\ac{MSE}} 
\begin{figure*}
    \centering
    \includegraphics[scale=0.6]{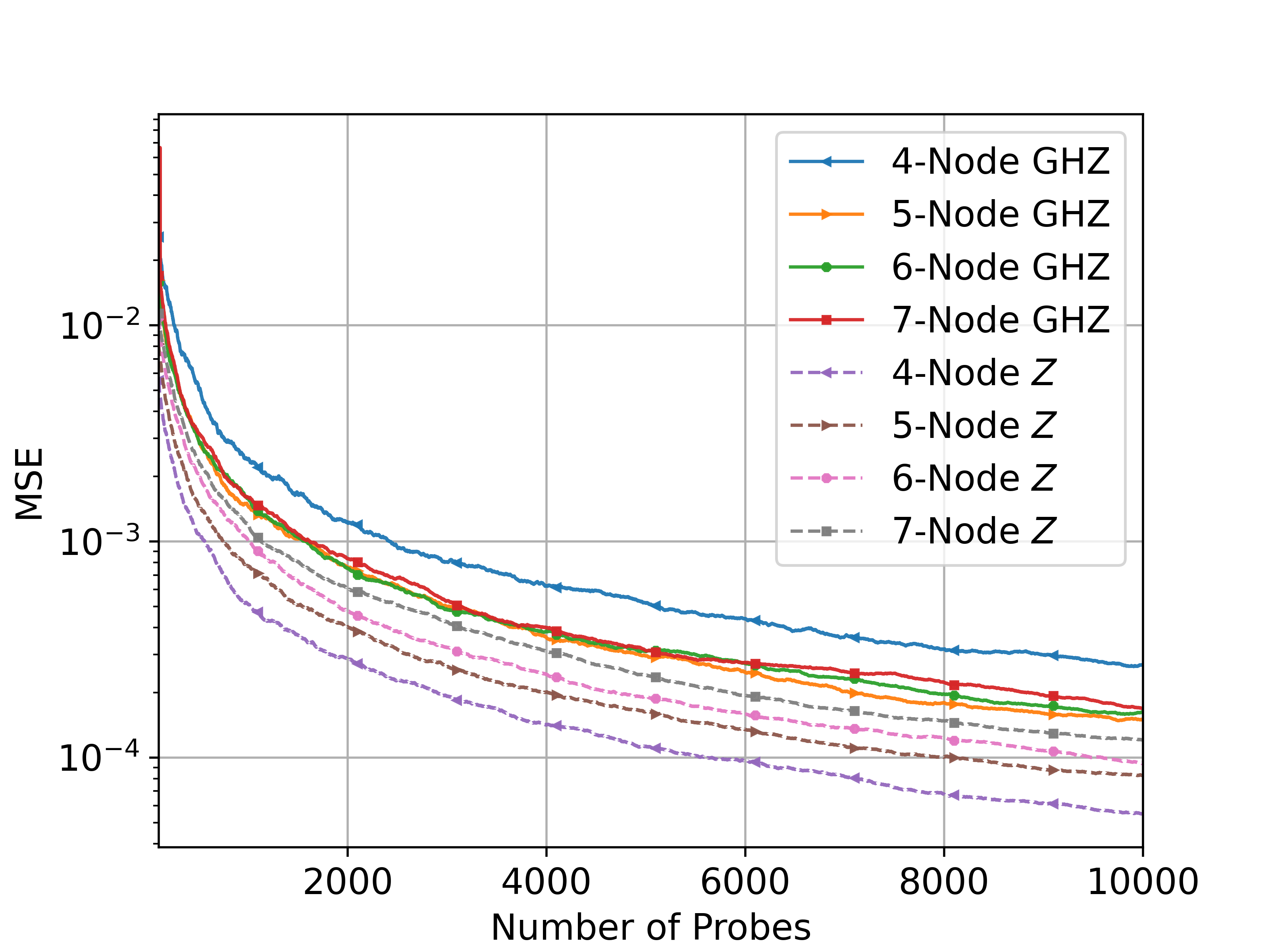}
    \caption{Mean squared error with number of samples. Simulations used stars with uniform depolarizing probability of $0.1$ per link. We vary the number of samples from $10^{2}$ to $\times10^{4}$. Markers are placed every $10^{3}$ samples. We average results on $200$ trials for each value of the depolarizing probability.}
    \label{fig:mse}
\end{figure*}

We now shift gears and analyze the \ac{MSE} of our estimators with respect to the number of samples used. Differently than the \ac{QCRB}, which only depends on the distributed state $\rho(\theta)$, the \ac{MSE} depends on the particular estimator obtained from measurements of $\rho(\theta)$. The \ac{MSE} provides insight on the number of samples needed for estimation in practical experiments, and the curves show the convergence rate of our estimators with the number of samples. Our results appear in Fig.~\ref{fig:mse} for star networks with uniform depolarizing probability of $0.1$ per link. Our results show that using thousands of samples for characterizing such noisy links yield an \ac{MSE} smaller than $10^{-2}$ for all network sizes considered and for both $Z$-diagonal and GHZ-diagonal states. Understanding how many samples are required for reliably characterizing links is of profound interest. This number can be used to compute a lower bound on estimation delay based on the state distribution rate with the Multicast circuit. If the end-to-end state distribution rate is on the order of KHz, the minimum delay expected to estimate parameters with an error on the order of $10^{-3}$ is on the order of seconds. If MHz distribution rates can be achieved, the latency lower bound for estimating parameters with the same confidence reduces to the order of milliseconds.

The MSE results also show that our estimators, which are based on bit-flip statistics of measurements, do not achieve the performance displayed by the QCRB. This is clear since the ordering of the curves for the MSE do not match that of the QCRB curves. Estimators based GHZ-diagonal states also require more samples to converge than estimators based on $Z$-diagonal states, given a fixed precision. Moreover, the MSE for estimators based on GHZ-diagonal states exhibit non-trivial behavior with network size, since there is no clear ordering of the curves with the number of end-nodes opposing to the separable state distribution case. This lack of ordering may originate from statistical fluctuations and we leave a more detailed exploration of this phenomena as theme for future work. Interestingly, this method yields identifiability for the entire parameter regime except when $p = 0.75$.

\section{Challenges and Open Problems}\label{sec:challenges}

Our assumptions and problem formulation allowed an initial investigation of end-to-end estimation in quantum networks. We now discuss research directions in advancing the \ac{QNT} knowledge frontier. We first discuss what relaxing the five assumptions described in Section~\ref{sec:assumptions} entails, and then present additional directions for future work.

\subsection{QNT with imperfect memory and operations}


Memory decoherence and quantum operation errors, e.g., non-unitary gates and imperfect measurements, are unavoidable in both near- and medium-term quantum hardware, having direct impact on the proposed QNT protocols. Thus, understanding how such errors modify the dependency of distributed states with link parameters is important to the application of our current \ac{QNT} protocols in practical scenarios. This dependency shapes the form of estimators and their performance, and operation noise and memory decoherence may require modifying our estimators. A thorough analysis of the effects of such errors to our protocols is theme for future work.

Moreover, one may also consider a more general version of the tomography problem where memory and operation errors must be estimated together with link errors in an end-to-end fashion. Developing protocols for this more general version of the \ac{QNT} problem is an exciting research direction.

\subsection{Identifiability of Arbitrary Pauli Channels}


Our methods are suited to the end-to-end characterization of bit-flip and depolarizing quantum channels, which are both described by one parameter. 
Adapting these methods to the characterization of arbitrary Pauli channels in star networks with parameter identifiability is one of our key near-term goals.
There are two key difficulties in characterizing Pauli channels with the protocols previously proposed. First, each channel requires the characterization of three parameters. Second, arbitrary Pauli channels modify the parameter dependence of the distributed states and previous estimators lose their meaning.
As previously discussed, characterizing Pauli channels is of profound importance due to their fundamental properties in quantum information theory. Applying the \ac{QNT} framework to their characterization remains an open problem and is the focus of ongoing work.

\subsection{QNT under restricted entanglement distribution services}
\ac{QNT} protocols discussed in this article assume that end-nodes can request the network to distribute quantum states with specific distribution circuits. This assumption requires classical information to be exchanged among nodes, detailing the gates that must be applied throughout state distribution. It is also of interest to investigate \ac{QNT} protocols that operate on distribution circuits directly provided by the network. For instance, it is likely that quantum networks will provide entanglement generation for end-users in the form of GHZ states---note that Bell states are a special case of GHZ states--- and a natural question is how to perform QNT utilizing such states directly without any additional quantum operations applied in the network nodes. QNT methods relying on such distribution services are less powerful than methods assuming that end-nodes can specify distribution circuits, although utilizing pre-defined distribution services reduces the amount of network cooperation required for estimation.

\subsection{Topology Estimation}

Our study on end-to-end estimation of quantum channels relied on the assumption that network topology is known to users, as captured by the \textbf{route selection} assumption. This assumption usually does not hold in a classical network, since users do not know the network topology. Classical network tomography has addressed this issue, investigating the problem of end-to-end topology estimation. Previous work has demonstrated topology characterization from end-to-end loss estimation in Multicast trees~\cite{anirudh2019occam}.
Thus, a clear research direction in \ac{QNT} is to investigate topology estimation through quantum state distribution. In particular, it is of interest to identify whether or not entanglement can provide advantages on the identification of network topology, and if Pauli channel estimation can be applied to topology estimation, following the findings for loss channels.

\subsection{Loss Estimation}

Loss is the main obstacle in fiber-based entanglement distribution. The tomography methods previously describe overcome this obstacle by using post-selected distributed quantum states. In essence, estimators rely on quantum states that have been successfully distributed in the network and the unique effect of loss is the increase in estimation delay. A natural research direction is to break ties with post-selection and investigate the estimation of loss through quantum methods. Loss estimation was extensively investigated in classical network tomography~\cite{he2021network} and searching for quantum advantages in the estimation of loss will shed light on the benefits of entanglement for end-to-end estimation. Note that loss can break the entanglement structure of distributed states, i.e., the loss of a qubit of a GHZ state leaves the remaining qubits in a maximally mixed state, which brings interesting caveats to estimation that require investigation in future work.

Utilizing multiparty state distribution for learning Pauli noise can be applied together with classical network estimation protocols for the estimation of loss under specific conditions. In particular, if network links can be modeled as a composition of loss and Pauli channels, multicast inference can be used together with Pauli estimators to simultaneously characterize both behaviors. Understanding the limitations of such simultaneous characterization is an interesting point for future work in \ac{QNT}.

\subsection{Adaptive \ac{QNT}}

As previously identified in~\cite{de2023characterization} for flip channels, estimator efficiency depends on parameter values. Thus, one should explore this dependency to define adaptive \ac{QNT} protocols that explore the different performance of estimators. A simple way to perform adaptive estimation within the~\ac{QNT} framework is to create a bank of \ac{QNT} protocols, evaluate their performance \textit{a priori}, and use this evaluation to decide which combination of state distribution circuits and measurements to utilize based on current link estimates. There is freedom on how to evaluate protocol performance, which can be done analytically, numerically, or through simulation. This simple approach serves as a starting point for the analysis of adaptive estimation, which can greatly benefit from Machine Learning algorithms.

Adaptive strategies are often required in multi-parameter quantum estimation in order to obtain estimators that satisfy the \ac{QCRB} with equality~\cite{liu2019quantum}. Here, adaptive  assumes a slightly different meaning: given a quantum state $\rho(\vec{\theta})$, the optimal measurement for estimation depends on $\vec{\theta}$, and the measurement basis must be steered dynamically to its optimal value. Therefore, arbitrary \ac{QNT} formulations may also require this form of adaptive estimation to attain the \ac{QCRB} performance.

\subsection{Characterizing Arbitrary Networks}\label{sec:arbitraryNets}

\begin{figure*}
    \centering
    \includegraphics[scale=0.57]{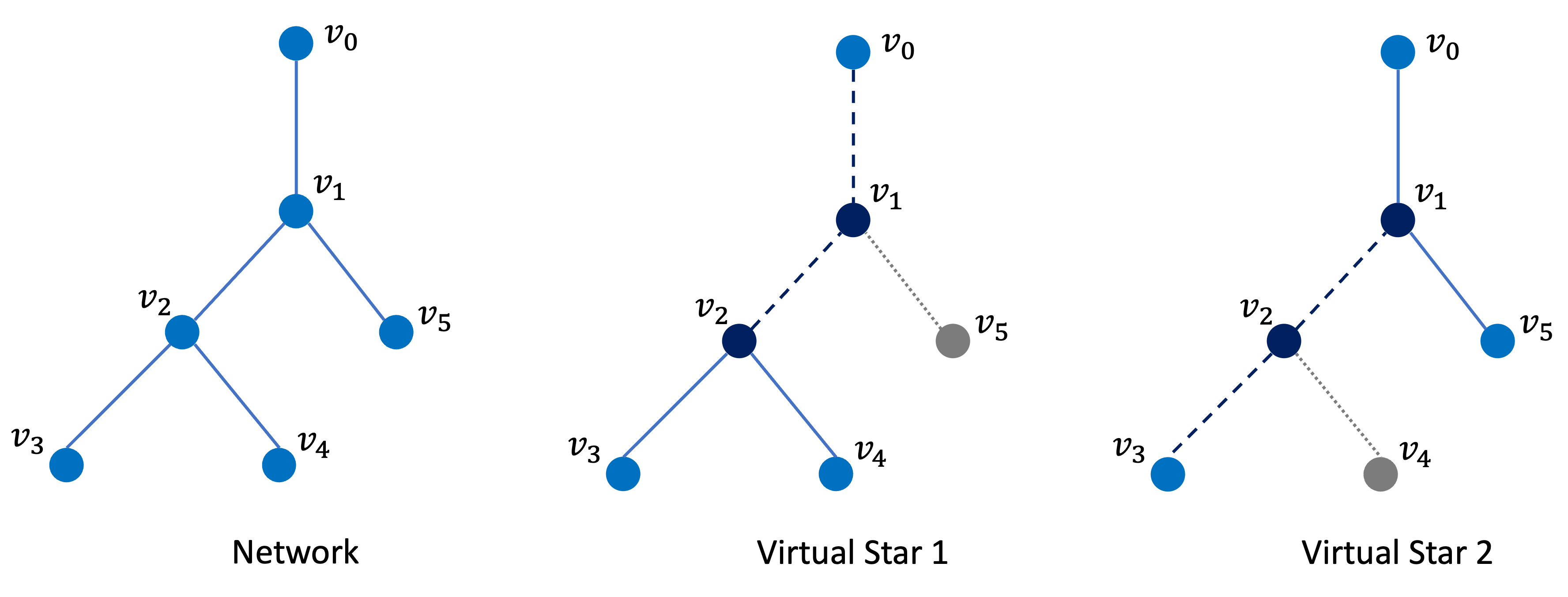}
    \caption{Decomposition of a tree into virtual stars. Paths $(v_0, v_1, v_2)$ and $(v_1, v_2, v_3)$ are virtual links in stars 1 and 2, respectively. Nodes $v_5$ and $v_4$ are not used in stars 1 and 2, respectively. Applying star characterization in virtual star 1 provides estimates for links $(v_2, v_3)$ and $(v_2, v_4)$ and for the virtual link $(v_0, v_1, v_2)$. In the case of virtual star 2, links $(v_0, v_1)$ and $(v_1, v_5)$ and the virtual link $(v_1, v_2, v_3)$ are characterized. Link $(v_1, v_2)$ can be estimated by triangulating parameter estimates from both stars.}
    \label{fig:decomposition}
\end{figure*}

Advancing the field of \ac{QNT} requires expanding the end-to-end estimation of stars to arbitrary quantum networks. The fundamental challenge in modifying protocols for stars to arbitrary topologies is the increase in complexity in estimating parameters end-to-end when paths interconnecting end nodes are longer than two links. As initially discussed in~\cite{de2022quantum}, one way to apply star characterization to identify parameters in more complex topologies is through star decomposition, which we depict in Fig.\ref{fig:decomposition}. A suitable star partition divides the network into virtual stars, i.e., stars whose links represent entire physical paths starting at end nodes. Virtual stars are characterized through the methods defined and an additional step is used to triangulate physical link parameters from virtual ones. A formal description of this method and a detailed analysis of its performance remain open.

One can also address \ac{QNT} of networks of arbitrary topologies without relying on star characterization. The \ac{QNT} framework is still relevant in this case since it has no assumptions on network topology. Finding a suitable set of quantum states and measurements for the framework is analogous to finding an informationally complete set of states and measurements in \ac{QPT}. This translates to the fundamental question of end-to-end parameter identifiability in quantum networks: under which conditions are link-noise parameters identifiable with end-to-end measurements? This question captures the fundamental limits of end-to-end noise characterization in quantum networks and remains an open problem.

\subsection{\ac{QNT} with Quantum Key Distribution}

Protocols for \ac{QKD}~\cite{bennett2014quantum} rely on classical messages exchanged between end-nodes to determine the security of distributed keys. Link noise corrupts states used for key distribution, and measurement statistics that specify if keys have been compromised depend on network noise. These measurement statistics can be useful in the \ac{QNT} context, and we point the application of \ac{QKD} protocols for \ac{QNT} as an exciting research direction for further investigation.

\subsection{\ac{QNT} with Syndrome Measurements}

\ac{QEC} is the ultimate way of dealing with the impacts of noise in quantum information. Error correction relies on error syndrome measurements, i.e. quantum non-destructive measurements, to identify when errors have occurred in quantum operations. Recent work has shown that syndrome measurement statistics can be used to estimate Pauli channels acting on encoded qubits~\cite{wagner2022pauli}. Therefore, a natural question to explore is the application of syndrome measurements for end-to-end characterization of quantum networks.

\section{Conclusions}

Quantum network tomography is a field of inquiry in its infancy that refers to the end-to-end characterization of quantum networks. In this article, we provided an introductory overview of initial \ac{QNT} results and discussed key assumptions and problem formulations. Furthermore, we advanced the state-of-the-art by defining estimators for depolarizing stars based on a previous \ac{QNT} protocol for characterizing flip stars.

QNT attempts to provide efficient (i.e., polynomial time in network size) end-to-end estimation of network noise parameters. We believe that \ac{QNT} will be instrumental in developing quantum networks by providing ways to verify and validate physical implementations. Moreover, we hope end-to-end estimation will bring powerful insights to network protocol design. It will enable ground-breaking applications that rely on noise characterization to improve efficiency and bring us closer to the revolution in communications promised by quantum networks.

{\em Acknowledgments}---This research was supported in part by the NSF grant CNS-1955744, NSF- ERC Center for Quantum Networks grant EEC-1941583, and MURI ARO Grant W911NF2110325.

\bibliography{ref}
\bibliographystyle{unsrt}

\end{document}